\documentclass[sigconf, noacm]{acmart}
\usepackage{stmaryrd}
\usepackage[utf8]{inputenc}
\usepackage{multirow}
\usepackage{multicol}
\usepackage{hyphenat}
\usepackage{amsmath}
\usepackage[noend]{algpseudocode}
\usepackage{algorithmicx,algorithm}

\usepackage[dvipsnames]{xcolor}
\usepackage{booktabs}

\setcopyright{none}
\acmConference[Conference '26]{The International Conference on XXX}{January 1, 2026}{Zurich, CH}
\acmISBN{978-1-4503-XXXX-X/2026/01}

\begin{document}

\title{FPGA-Accelerated Lock Management and Transaction Processing: Architecture, Optimization, and Design Space Exploration}


\author{Shien Zhu}
\affiliation{%
  \institution{ETH Zurich}
  \city{Zurich}
  \country{Switzerland}}
\email{shien.zhu@inf.ethz.ch}
\orcid{1234-5678-9012}

\author{Gustavo Alonso}
\affiliation{%
  \institution{ETH Zurich}
  \city{Zurich}
  \country{Switzerland}}
\email{alonso@inf.ethz.ch}
\orcid{1234-5678-9012}

\renewcommand{\shortauthors}{Shien Zhu et al.}

\begin{abstract}

Managing data consistency is one of the main bottlenecks in traditional transaction processing. Acquiring and releasing the many data locks needed by queries and transactions has a long latency on CPUs, especially when there is contention on hot locks. In addition, reading and writing the data creates another bottleneck in the memory access, limiting throughput. In this paper, we explore the possibility of offloading lock management and concurrency control to an accelerator to alleviate both bottlenecks. 
We do this through an efficient FPGA-based transaction processing accelerator with strong data consistency and good scaling capability. First, we propose a fully-functional Lock Agent with integrated hardware data structures to store the lock status and thus reduce the lock serving latency. Second, we propose an asynchronous pipelined Transaction Agent to execute multiple transactions simultaneously. Third, we scale up the architecture through a channel-table hierarchy to further improve transaction and lock serving parallelism. Finally, we conduct a design space exploration to tune the hyper-parameters and gain insights. Experimental results show that the accelerator achieves 6.3-16.7 million lock/s using 4 lock agents. The transaction processing throughput is 60.4-70.9 thousand transaction/s/agent on the TPC-C benchmark, which is 38.6-50.9$\times$ faster than the CPU baseline. The overall performance is 4.8-6.5$\times$ higher than the CPU baseline without lock table size limitation. 



\end{abstract}


\begin{teaserfigure}
  \includegraphics[width=0.8\textwidth]{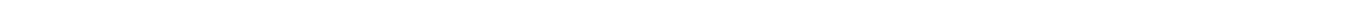}
\end{teaserfigure}


\maketitle

\section{Introduction}

Online Transaction Processing (OLTP) is an essential use case of database systems \cite{TOCS_2013_Spanner, ICDE_2022_PolarDB}. To achieve the Atomicity, Concurrency, Isolation, and Durability (ACID) requirements of transactions, databases usually apply strict concurrency control policies, among which Multi-Version Concurrency Control (MVCC) \cite{SIGMOD_2015_fast-MVCC, VLDB_2017_In-Mem-MVCC} and 2-Phase Locking (2PL) \cite{TODS_1993_2PL, PPPP_2023_2PL-StarvationFree} are the most commonly used. The former is considered to be faster as it requires less locks, but the latter is considered to be stronger from a consistency point of view.  

Especially with 2PL, lock management is one of the main bottlenecks \cite{VLDB14-Abyss, VLDB_2015_VLL-Lock-Manager, VLDB_2018_Lock-Scheduling}. Locks need to be acquired for every tuple accessed, which is an expensive operation causing contention, and that becomes a serious limitation to scalability in distributed settings due to the overhead of acquiring remote locks \cite{FAST23-RangeLocks, SIGMOD-18-Distributed-locking}. 
Recent work has explored the use of Remote Direct Memory Access (RDMA) to reduce communication latency for locking in distributed settings \cite{VLDB17-End-of-Myth,FAST23-RangeLocks,SIGMOD-18-Distributed-locking}, but the underlying overhead of lock management remains if strong consistency is the goal \cite{VLDB_2019_2PhaseLocking} . 

In this paper, we explore the possibility of offloading the lock management and concurrency control operations to an accelerator instead of running them on a CPU as it is commonly done in both commercial products and in research. Based on the observations made in \cite{VLDB_2019_2PhaseLocking} indicating that 2PL works well in modern database systems and the approach provides strong consistency, we focus on implementing 2PL, which is the most demanding of the approaches to implement concurrency control. To prototype the accelerator, we use FPGAs to take advantage of their ability to process data at line rate and the availability of High-Bandwidth Memory (HBM) for high-bandwidth data access. FPGAs are already used in database engines \cite{VLDB_2020_IMDB-FPGA} including commercial ones \cite{SIGMOD19-X-Engine,FPGA-PolarDB-FAST20}, although not for concurrency control,to implement, e.g., joins \cite{ISCAS_2025_Range-Join}, filtering \cite{Ibex13,YourSQL16,VLDB19-Accorda,ICDE_2022_PolarDB}, compression and encryption \cite{Hana-FPGA22}, and data structure compaction \cite{SIGMOD19-X-Engine}, and SQL2FPGA \cite{TRETS_2024_SQL2FPGA} further provides an automated compiling framework to support hybrid CPU-FPGA execution.

\begin{figure}[t]
    \centering
    \includegraphics[width=0.85\linewidth]{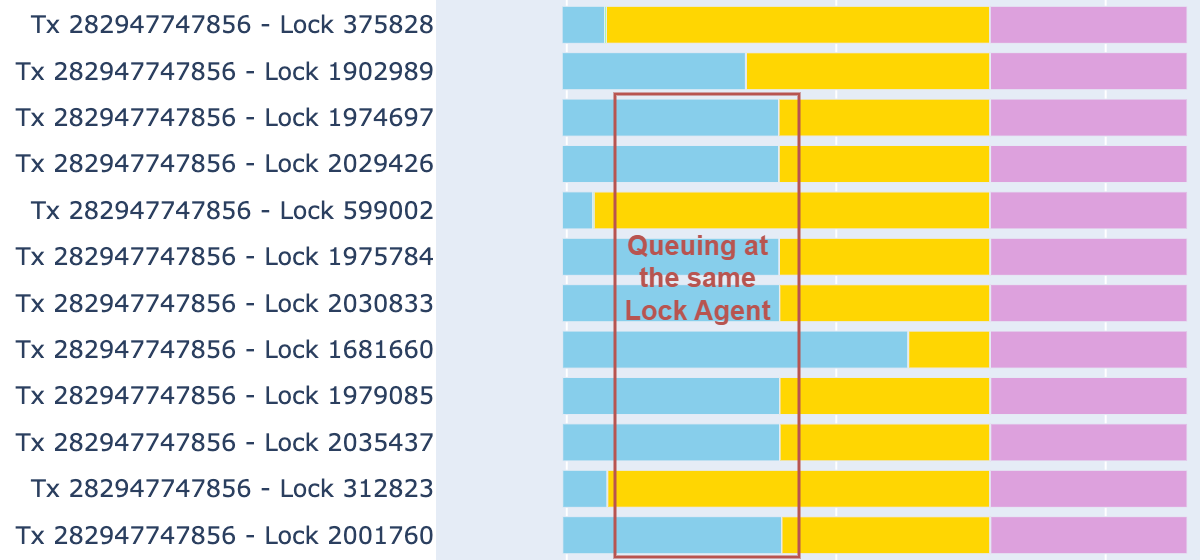}
    \caption{Part of a profiled TPC-C transaction's timeline. These 7 lock requests are waiting for the same lock agent.}
    \label{fig-cpu-queuing}
\end{figure}

Designing an OLTP accelerator on FPGA is not trivial. There are three key challenges. First, how to minimize the latency and maximize the throughput of lock serving. As Figure \ref{fig-cpu-queuing} shows, profiling on CPU-based lock agents shows that locks suffer from long waiting latency before the lock is preprocessed by the lock agent. The average waiting time per lock agent is around 0.4 ms. The main reason is due to the long memory latency for the lock agent to access the lock status (usually implemented by hash tables and linked lists) to make decisions. As only a few locks in one transaction are hot locks which are frequently accessed by other transactions, the huge portion of cold locks may incur high cache misses and long memory access operations. Therefore, separation of the processing unit and the lock status causes a long waiting latency. Second, how to maximize transaction execution efficiency and parallelism under the hardware resource constraints of FPGA. As each transaction agent has several execution stages with long time spans, simply executing one transaction per agent is inefficient. In addition, due to the complexity of a transaction agent, simply increasing the number of agents for higher parallelism will run out of FPGA resources. Third, how to balance the lock serving parallelism and the crossbar size between transaction agents and lock agents. As the lock may reside in any lock agent, a crossbar is needed to connect the transaction agents and the lock agents. However, large crossbar sizes consume more hardware resources and may bring a high failure rate to the placement and routing stage.

To address these challenges, we have developed a highly parallel transaction processing accelerator with strong data consistency on FPGAs. First, we integrate the hardware lock tables into the lock agents to reduce the lock serving latency, utilizing the fast on-chip memory of FPGA to build hash tables and linked lists. Second, we design an asynchronous pipelined Transaction Agent to decompose the life cycle of one transaction into several stages, and execute multiple transactions on one agent simultaneously using context switching to fill the pipeline and overlap the lock serving latency between transactions. Third, we scale up the architecture with multiple transaction and lock agents to increase the parallelism. Observing that the lock serving usually takes 3-20 clock cycles, we apply a lock channel-table hierarchy to reduce the crossbar size to balance the parallelism and resource utilization. Finally, we conduct a design space exploration to study the lock serving efficiency of different lock channel-agent combinations and the efficiency of transactions by context switch per agent. 

The accelerator achieves 6.3-16.7 million lock/s using 4 lock agents. The transaction processing throughput is 60.4-70.9 thousand transaction/s/agent on the TPC-C benchmark, which is 38.6-50.9$\times$ faster than the CPU baseline under the same configuration. The overall performance of our accelerator is 4.8-6.5$\times$ higher than the CPU baseline without lock table size limitation.   


\section{Background and Related Work} 

\subsection{Background}

The canonical approach to concurrency control in databases is 2 Phase Locking (2PL). In 2PL, transactions are allowed to acquire locks in two phases \cite{BernsteinHG87}. In the first phase locks can be acquired and, in the second phase, locks can only be released but no longer acquired. In practice and for recovery reasons, locks are usually kept until the end of a transaction and only then are released. The approach guaranties serializability, the standard correctness criterion for concurrency control, or, in other words, that the history produced is equivalent to a history where transactions are executed serially. Locks need to be obtained for all operations that might potentially conflict with each other: reads and writes on tuples.

The conventional way to implement database locking is using a hash table \cite{BlackBook92}. The id of a tuple that needs to be locked is hashed to find the corresponding bucket. In that bucket, there is a linked list of lock requests that is traversed to place the new request. A scheduler uses this list to decide when a lock can be granted depending on the compatibility across lock types and what locks are being held at the moment on a given tuple. The design of the hash table plays a big role in the performance of locking, as, depending on its size, it might induce false conflicts and introduce additional latency when having to navigate long linked lists.  Our implementation uses a similar approach and data structure to make it comparable to existing designs and compatible with deployed transaction managers. 


This overhead has led to alternative designs that try to avoid or minimize locking and coordination, either through algorithmic approaches such as optimistic concurrency control or timestamp ordering, or by relaxing the correctness criteria. Comparison of these protocols and approaches has been extensively studied in the literature (e.g. \cite{VLDB14-Abyss}),  but strong consistency is the preferred option if it can be implemented with sufficient performance \cite{VLDB_2019_2PhaseLocking}.

\subsection{Related Work}

Transaction processing is the crucial performance component of OLTP systems. As such, it has been extensively studied for almost 4 decades both from a formal/theoretical perspective as well as in terms of efficient implementations \cite{BernsteinHG87,BlackBook92,CritiqueSQL95,VLDB14-Abyss,VLDB_2019_2PhaseLocking}. Because it is such a critical component and an inherent bottleneck in case of conflicting accesses, many approaches have been proposed over the years. In this paper, we focus on the conventional 2PL design as it is the most complete and challenging. Many other approaches can be implemented as variations or optimizations of this basic design. 

FPGA acceleration for database applications has emerged as a significant research area in the last decade \cite{VLDB_2020_IMDB-FPGA, ParaComp_2024_FPGA-Relational-DB, CompSurvey_2023_Non-RelationalDB-FPGA}. \cite{PACT_2012_FPGA-IMDB-Decompress} offloads the data decompression and predicate evaluation operations from CPU to FPGA accelerator through PCIe with software compatibility. \cite{SIGMOD19-X-Engine} implements LSM-tree compaction for Alibaba's cloud database. \cite{Ibex13,YourSQL16,FPGA-PolarDB-FAST20} integrate an FPGA with the storage engine to accelerate data filtering and reduce CPU load. \cite{Sketches23} proposes an FPGA accelerator implementing sketches over large data collections. \cite{Hana-FPGA22} uses an FPGA to implement compression and encryption on the I/O path for SAP Hana. \cite{TRTS_2025_FPGA_Delta_Merge} accelerates the time-consuming index recode of the delta merge process in in-memory databases using an FPGA programmed with OpenCL and High-Level Synthesis (HLS) to improve OLTP throughput. It implements columnar storage with greater compression rates than row-based ones, which also benefits OLAP applications. SQL2FPGA \cite{TRETS_2024_SQL2FPGA} provides an automated compiling framework to support hybrid CPU-FPGA execution using the optimized query plans from database engines. In summary, most of these FPGA accelerators for databases focus on the Online Analytical Processing (OLAP) scenario with heavy computation. 





\section{Transaction Processing System}

\subsection{System Architecture}
\begin{figure}[t]
    \centering
    \includegraphics[width=\linewidth]{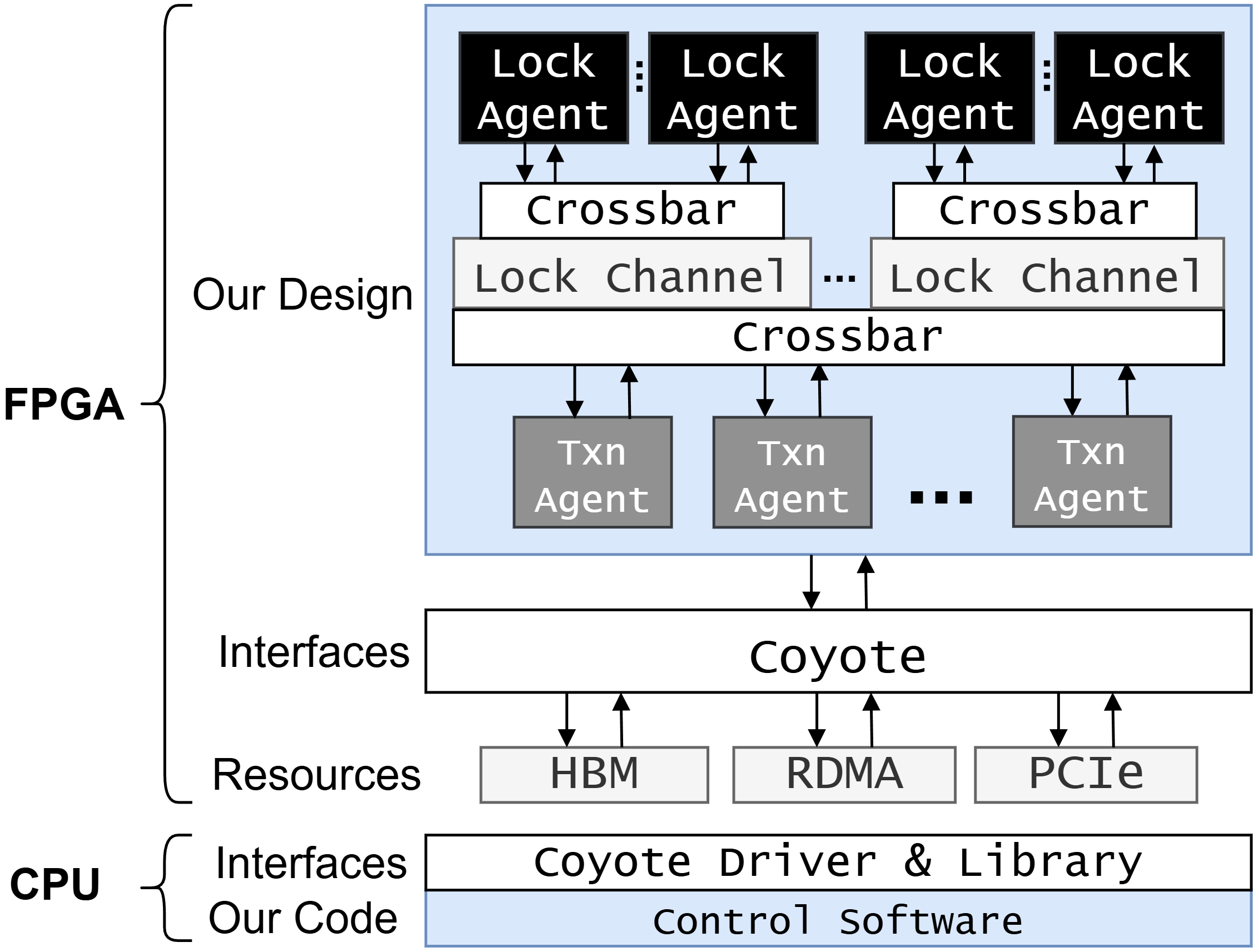}
    \caption{The FPGA-based transaction processing system.}
    \label{fig-system-architecture}
\end{figure}

Based on the 2PL concurrency control policy, we propose an FPGA accelerator for OLTP that supports the whole workflow of lock serving and transaction processing. The architecture is optimized for high parallelism on transactions and locks, low lock serving latency, good scalability, and reasonable resource utilization. As Figure \ref{fig-system-architecture} shows, the FPGA-based OLTP system contains both software and hardware components highlighted in blue. 

\textbf{Accelerator}: Our accelerator includes dedicated Transaction Agents and Lock Agents to improve the OLTP parallelism and reduce the lock serving latency. The Transaction Agent is responsible for the entire transaction workflow, including loading transactions, requesting and releasing locks, and committing transactions. The Lock Agent accepts lock requests, decides whether to grant the lock, and sends lock responses. It has a built-in hash table as the lock table and a linked list to store the locks that conflict with the current lock status but may be granted later. The Transaction Agents and Lock Agents are connected via two levels of crossbar to reduce the interconnect complexity. We cluster multiple Lock Agents into one Lock Channel, then connect the Lock Channels to the Transaction Agents via an all-to-all crossbar. 

\textbf{Interface}: Based on Coyote \cite{Coyote_OSDI_2020, CoyoteV2_SOSP_2025}, our FPGA accelerator can access the on-board resources and communicate with the control software on the CPU. The Coyote serves as the interface to the HBM on the FPGA board and the PCIe that connects to the CPU. In addition, Coyote provides the Linux driver and a library to interact with its interfaces. Adopting Coyote as the FPGA shell greatly reduces the engineering complexity and provides future extension possibility of multi-node processing using the RDMA network stack. 

\textbf{Control Software}: The control software has two main functions: transferring the transaction workload and sending control signals. The transaction workload is initialized on the CPU and then sent to the FPGA's HBM. We control the accelerator behavior by configuring the FPGA hardware Control Signal Registers (CSRs). This involves specifying the memory addresses of the transaction workload and the database for transaction commits, setting the start signals to trigger transactions, and reading CSRs to monitor the transaction progress and relevant statistics. 

\subsection{Lock Agent}
The Lock Agent serves locks to all transaction agents that query it. Lock Agents accept Get and Release lock requests, make decisions based on the lock table states, and send the lock responses in one of these states: Granted, Released, Waiting, and Aborted. We optimize Lock Agents for low latency to reduce congestion and prevent long request queuing before entering the agents.

\subsubsection{Architecture} 
\begin{figure}[t]
    \centering
    \includegraphics[width=0.9\linewidth]{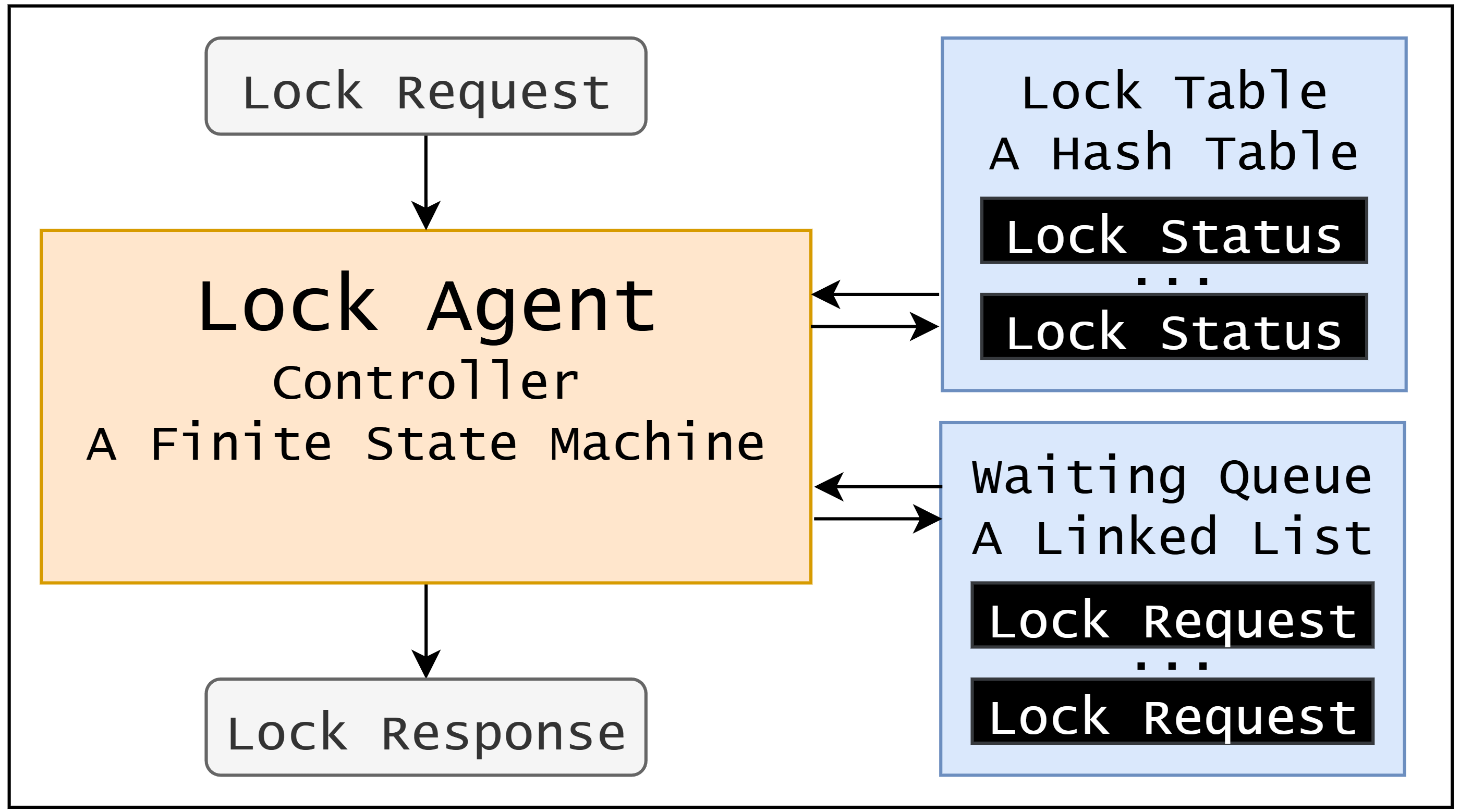}
    \caption{Lock Agent architecture.}
    \label{fig-lock-agent}
\end{figure}

$\\$
Figure \ref{fig-lock-agent} presents the hardware architecture of our proposed Lock Agent. The Lock Agent has a simple Input/Output (I/O) interface, comprising the input lock request and the output lock response. The Lock Agent controller is a complex Finite State Machine (FSM) that processes both Get and Release lock requests, which will be presented in detail in Section \ref{section-lock-serving-policy} Lock Serving Policy. 

We decouple the lock table that stores lock information as a hash table and a linked list. The \textbf{hash table} stores the lock statuses for quick reference. The hash table entry contains an implicit lock identifier (lockID), the lock mode (referring to Table \ref{tab:lock-modes} for more details), the number of owners of this lock, the address of the waiting queue, and valid signals of stored information. The l\textbf{inked list} stores requests for locks that conflict with the current lock status, and these locks can be granted once other transactions release them. If a Get lock request conflicts with the current lock mode, we will put it to the waiting queue for quick grant when the current lock is released later. The linked list entry contains the Lock Request in waiting status, the next entry address of the same lock, and valid signals of the stored information. 

To achieve ultimate low latency, the control logic of the hash table and linked list are integrated into the FSM controller. A standard hardware hash table has Insert, Delete, Update, and Query operations, but we keep the data structure of the hash table and fuse the control logic to the FSM. Similarly, a standard hardware linked list has Insert, Delete, and Pop operations, and we fuse its logic to the FSM too. As a result, we reduce the clock cycles that send commands and receive results from these hardware data structures, and only the processing time counts in the critical path.

\begin{figure}[t]
    \centering
    \includegraphics[width=0.9\linewidth]{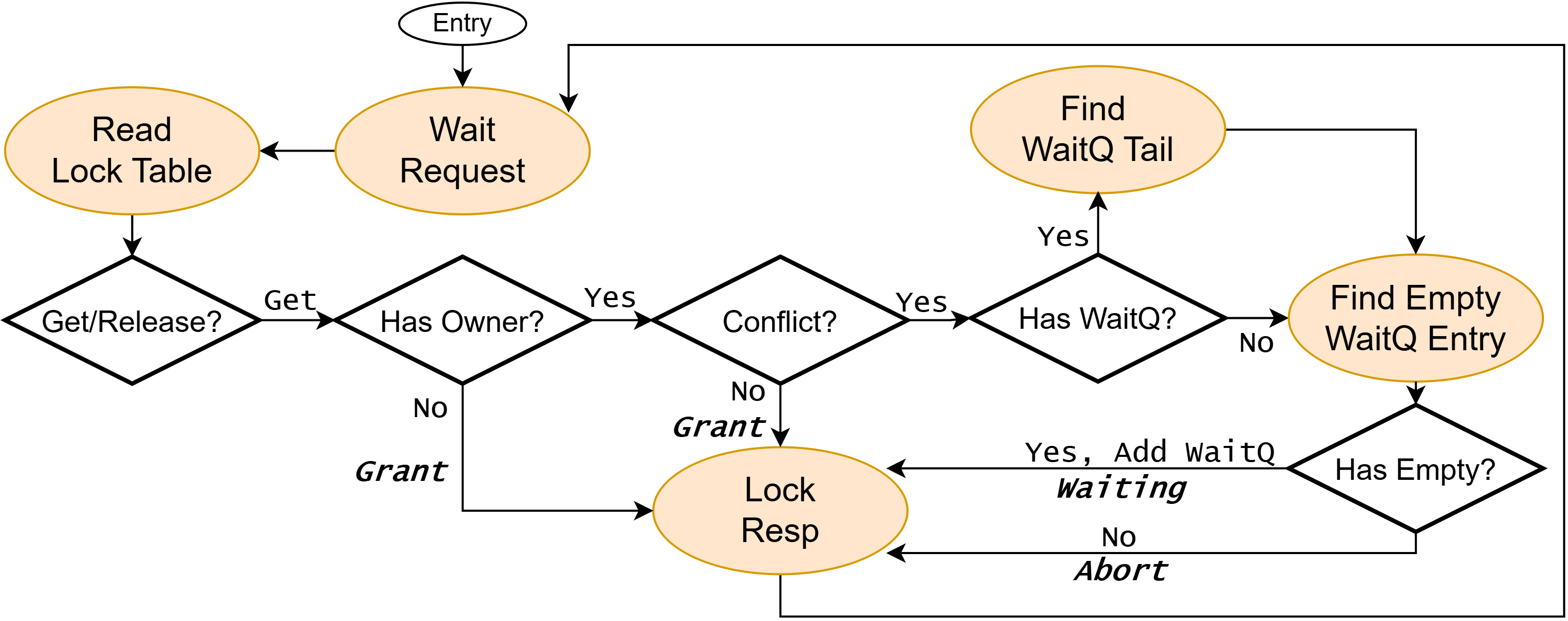}
    \caption{Control logic of Get lock requests.}
    \label{fig-lock-agent-get}
\end{figure}

\subsubsection{Lock Serving Policy} $\\$
\label{section-lock-serving-policy}
The lock serving policy is based on the lock mode compatibility of the lock requests and the lock status. To make it clear, we present the serving logic of Get and Release locks, respectively. 

\textbf{Lock Mode:} Following related work \cite{VLDB_1975_SIX-Locks, VLDB_2019_2PhaseLocking} and existing practice \cite{BlackBook92}, our lock has 6 modes expressed by three signals: Shared (S), Intent (I), and Exclusive (X). As Table \ref{tab:lock-modes} shows, the lock modes are No Lock (NL, SIX=000), Intent Shared Lock (IS, SIX=110), Intent Exclusive Lock (IX, SIX=011), Shared Lock (S, SIX=100), Shared and Intent Exclusive (SIX, SIX=111), and Exclusive Lock (X, SIX=001). The S and SIX locks trigger data read and the X locks trigger data write in the transaction commit stage. 

Table \ref{tab:lock-modes} shows whether two lock modes are compatible. The granted mode is the current lock status in the hash table, while the requested mode is asked by the input lock request. For example, if the corresponding lock table entry has no lock, then the Lock Agent can grant all types of lock requests. If the current lock status is S, then it can only grant the lock to lock requests in NL, IS, and S modes. Nevertheless, the NL lock does not appear in lock requests in real world applications. 

\begin{table}[b]
\centering
\caption{Lock request mode compatibility matrix \cite{VLDB_1975_SIX-Locks}. 
A checkmark (\checkmark) indicates that the incoming requested mode (row) is compatible with the current granted mode (column).}
\label{tab:lock-modes}
\begin{tabular}{lcccccc}
\toprule
\textbf{Requested} & \multicolumn{6}{c}{\textbf{Granted Mode}} \\
\cmidrule(lr){2-7}
\textbf{Mode} & NL & IS & IX & S & SIX & X \\
\midrule
NL  & \checkmark & \checkmark & \checkmark & \checkmark & \checkmark & \checkmark \\
IS  & \checkmark & \checkmark & \checkmark & \checkmark & \checkmark &   \\
IX  & \checkmark & \checkmark & \checkmark &   &   &   \\
S   & \checkmark & \checkmark &   & \checkmark &   &   \\
SIX & \checkmark & \checkmark &   &   &   &   \\
X   & \checkmark &   &   &   &   &   \\
\bottomrule
\end{tabular}
\end{table}

\textbf{Get Locks:} Figure \ref{fig-lock-agent-get} presents the part of the FSM controller for Get lock requests. The Lock Agent stays in the Wait Request state by default. When a new lock request comes, it reads the hash table and makes decisions based on the lock status. First, it judges whether the lock request type is Get or Release. For the Get requests, it checks whether this lock has been granted before. If the lock has no owner, then we can Grant the lock immediately. Otherwise, the lock agent checks whether the lock conflicts with the existing grant mode according to Table \ref{tab:lock-modes}. If there is no conflict, then we can Grant the lock too. Otherwise, we need to insert this lock request into the waiting queue (waitQ) for future Grant response. If the requested lock has a waitQ, we find the tail of the waitQ and then find an empty entry to store the current lock request. If an empty waitQ entry is found, then we insert the lock request into the waitQ and respond with a Waiting signal. But if the empty entry cannot be found within certain clock cycles, we will send an Abort lock response. The number of waitQ entries visited for finding an empty entry is configurable. We set it to 8 or 16 in most cases to balance the success rate and latency of finding an empty waitQ entry. 

\begin{figure}[t]
    \centering
    \includegraphics[width=0.9\linewidth]{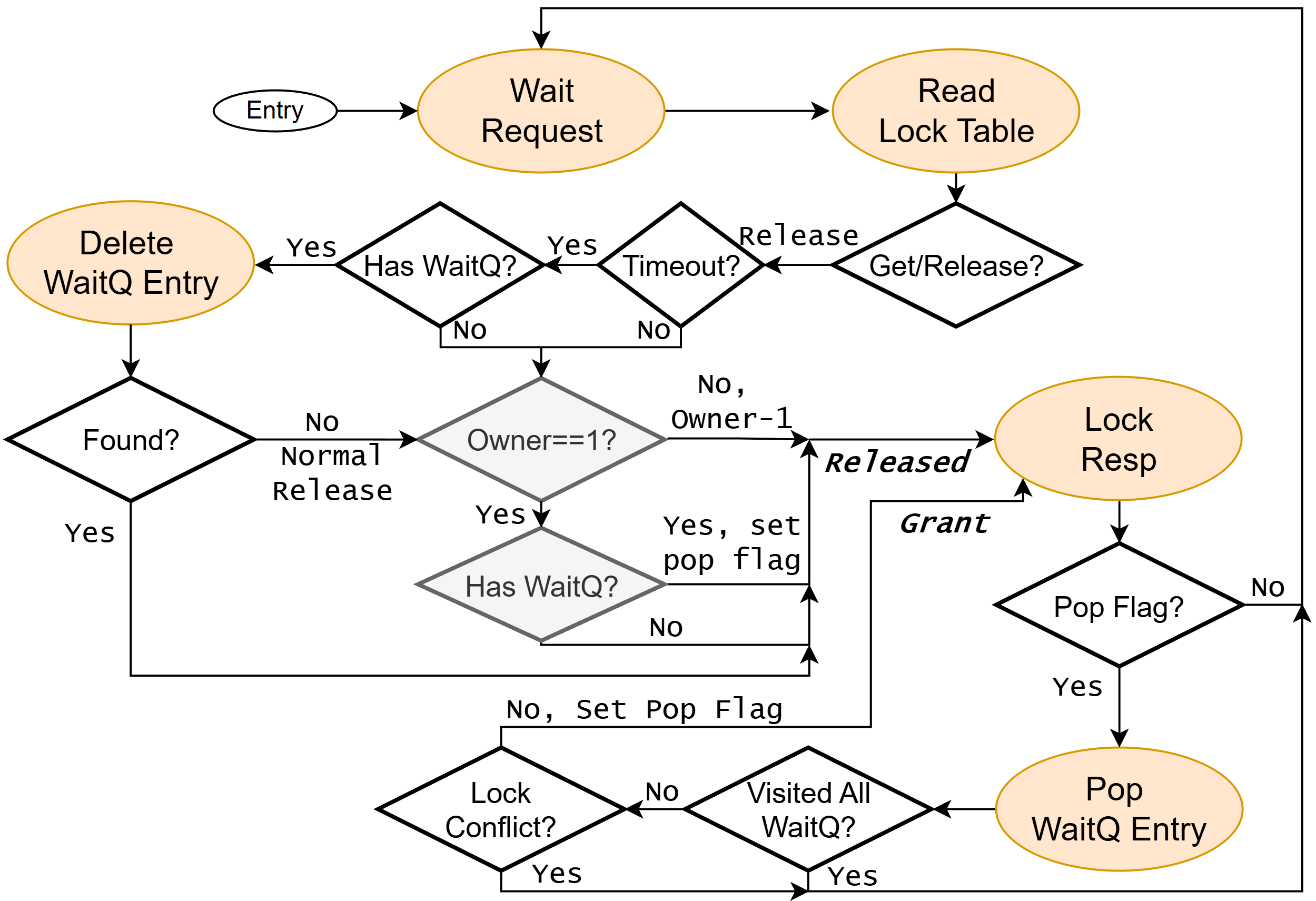}
    \caption{Control logic of Release lock requests.}
    \label{fig-lock-agent-release}
\end{figure}

\textbf{Release Locks:} The control logic of the Release lock requests is shown in Figure \ref{fig-lock-agent-release}. The states of Wait Request, Read Lock Table, and judgment on the lock request type are shared with the control logic of Get lock requests. When a Release request comes, the Lock Agent checks whether the release is due to timeout, indicating that the lock to release may be still inside the waitQ and not granted yet. The Lock Agent enters the Delete WaitQ Entry state when the waitQ is valid, otherwise it goes to the normal release checks. If the lock to release is found in the waitQ, then the Lock Agent sends the Released lock response. Otherwise, it will also go to the normal release checks. In normal release, if the owner of this lock is greater than 1, then the Lock Agent reduces the owner counter by 1 and sends the Released lock response. If there is only 1 owner and the lock has no waitQ, this lock entry will be cleared to No Lock and a Released lock response will be sent. 

\begin{figure*}[!h]
    \centering
    \includegraphics[width=0.9\linewidth]{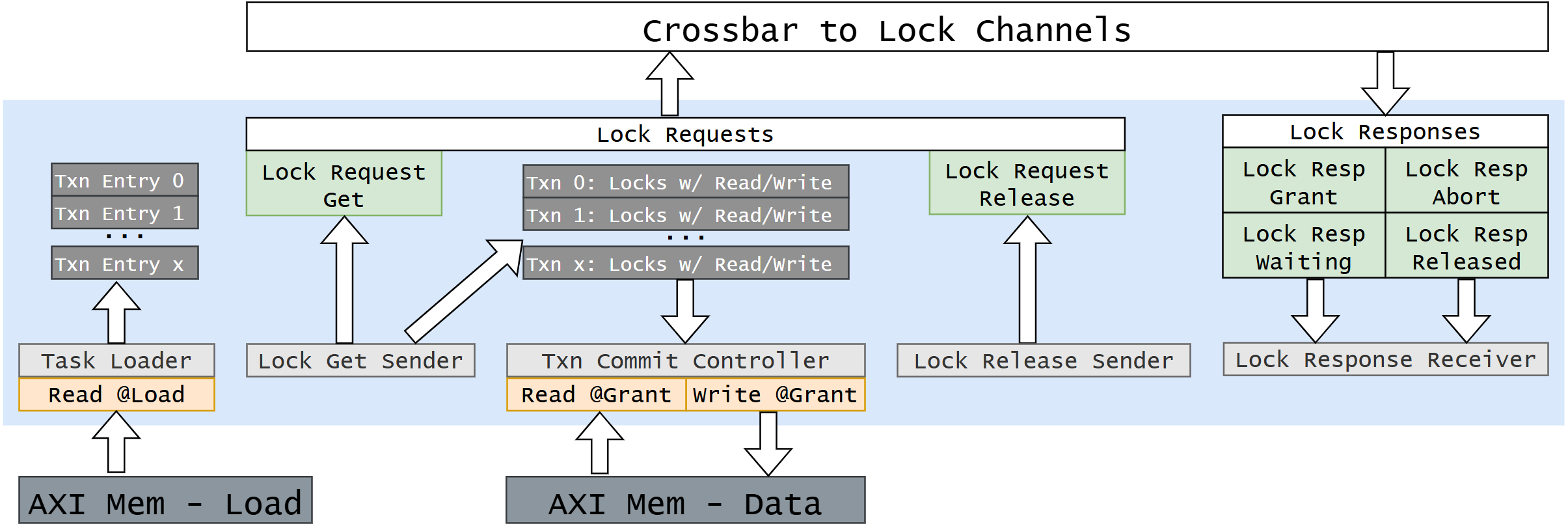}
    \caption{Transaction Agent architecture.}
    \label{fig-transaction-agent}
\end{figure*}

However, if the only 1 owner releases this lock and the waitQ is valid, we need to pop the waitQ. The Lock Agent will set a Pop Flag, release this lock with a Released response, clear the lock status to No Lock, and then move from the Lock Resp state to Pop WaitQ Entry. The first entry in the waitQ will always be granted because the No Lock mode is compatible with all lock requested modes. The Lock Agent updates the lock mode in the hash table, updates the Pop Flag, sends a Grant lock response for this waitQ entry, and deletes this waitQ entry. After sending the Grant response, the Lock Agent repeats the workflow to check the Pop Flag to pop more waitQ entries or go back to Wait Request state. It will consecutively grant compatible waitQ lock requests from the waitQ head until all waitQ entries of this lock are visited or a lock conflict occurs. This may better preserve the sequence between transactions, but may have a small performance penalty compared to popping all compatible lock requests in the waitQ. 

\subsubsection{Latency Analysis} $\\$
We analyze the lock serving latency to show that the Lock Agent is very efficient. As Figure \ref{fig-lock-agent-get} and \ref{fig-lock-agent-release} show, the best case latency of Get and Release lock requests are 3 clock cycles. For the Get request, all Grant responses can be sent in 3 clock cycles. The Wait Request state issues the hash table read directly when the new lock request arrives, and the Read Lock Table state gets the lock status in the next cycle to make decisions, then the lock responses are sent out in the third clock cycle. 

The incompatible lock modes of granted lock and requested lock lead to longer latency than Grant responses. The Waiting response usually takes 5-20 cycles (one cycle for updating the waitQ entry, longest waiting queue is 8, and the maximum waitQ entry searched is also 8), which increases with the length of the waiting queue and the availability of waitQ entries. Similarly, the Abort response takes 11-19 cycles because it must reach the upper bound of finding an empty waitQ entry. 

For the Release lock requests, the normal lock release (no timeout) also takes only 3 clock cycles. As Figure \ref{fig-lock-agent-release} shows, the Released responses are always sent first before going to the Pop WaitQ Entry state.  The timeout Release requests usually take 5-11 clock cycles (longest waiting queue is 8). It takes at least 1 cycle to find the waitQ entry, 1 cycle to delete the waitQ entry, and 3 cycles for the Wait Request, Read Lock Table, and Lock Resp states. The updating of relevant waitQ entries is fused to the Lock Resp state to reduce latency. After sending the Released lock response, each Grant response from the waitQ entry takes 3 cycles (Pop WaitQ Entry, delete the waitQ entry, and send the lock response). 


Taking the aforementioned configuration with 200MHz FPGA clock frequency as an example, the worst case of lock response in 20 cycles corresponds to 10 million served locks per Lock Agent. Such high lock serving efficiency allows us to explore some hyper-parameters to make a trade-off between the lock serving latency and the waiting queue length. On the other hand, the best case of Get and Release lock requests in 3 clock cycles should be preserved when we cluster the Lock Agents into Lock Channels. Thus, the maximum number of lock tables per channel should be no more than 4 to ensure that getting shared locks and normal lock release can still reach their lowest latency under saturated lock channels.

\subsection{Transaction Agent}

We design an asynchronous pipelined Transaction Agent by splitting the transaction processing stages. It can execute multiple transactions by context switch to fill the pipelines. The Transaction Agent is asynchronous because each component can work on different transaction as long as that transaction is ready for that processing stage. Therefore, we will introduce the transaction processing life-cycle, then present the detailed Transaction Agent architecture.

\subsubsection{Transaction (Txn) Processing Life-cycle} $\\$
We adopt the standard 2PL policy and adapt it for better efficiency on FPGA. Therefore, the transaction processing workflow is slightly different from related works. The FPGA-based transaction processing workflow is as follows:
\begin{itemize} 
    \item \textbf{Loading}: The txn context is loaded from on-board HBM to on-chip memory. 
    \item \textbf{Sending Get Requests}: The Txn Agent acquires all locks of this transaction. 
    \item \textbf{Obtaining Grant Responses}: The Txn Agent waits for lock grant responses. When it holds all locks, it will proceed to the Commitment stage. 
    \item \textbf{Committing the Txn}: The Txn Agent conducts all corresponding data read and write operations.
    \item \textbf{Sending Release Requests}: The Txn Agent sends lock requests to release the locks after the commitment stage.
    \item \textbf{Obtaining Released Responses}: The Txn Agent waits for lock released responses. When all locks have been released, it will proceed to the cleanup stage.
    \item \textbf{Cleanup}: The txn context is cleared, and the Txn Agent can start loading the next transaction. 
\end{itemize}

\begin{figure*}[]
    \centering
    \includegraphics[width=0.9\linewidth]{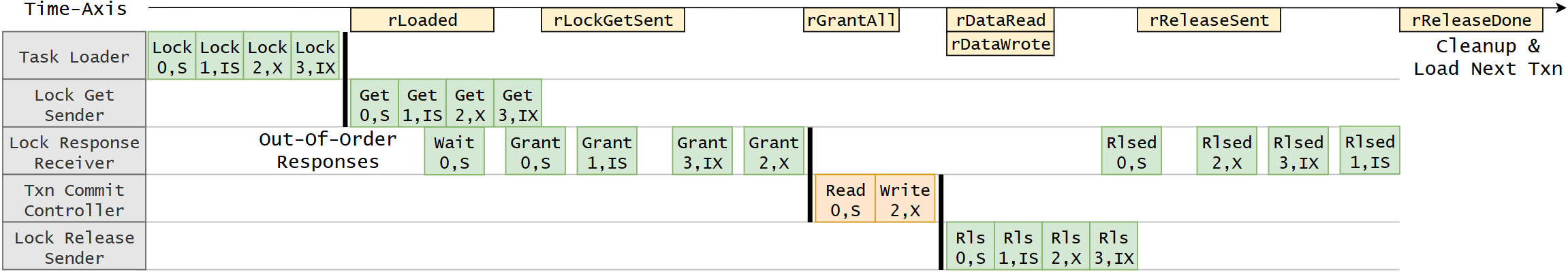}
    \caption{Pipeline of executing one example transaction.}
    \label{fig-one-transaction-pipeline}
\end{figure*}

However, a transaction may be aborted in certain cases. For example, one or more of the requested locks are denied by the Lock Agent and the Txn Agent receives Abort lock responses.  Or that txn does not receive all the lock granted responses before the timeout limit, then a timer will trigger the aborting workflow. When the abort signal is triggered, the Txn Agent will skip the commitment stage to send the Release lock requests directly. The Release lock requests have deterministic Released lock responses. Thus, there is no need to set a timer for the Released responses. The processing workflow for the aborted transactions is as follows:

\begin{itemize} 
    \item \textbf{Loading}  
    \item \textbf{Sending Get Requests} 
    \item \textbf{Obtaining Lock Responses - Abort/Timeout Signal Triggered}: Skip the commitment stage. 
    \item \textbf{Sending Release Requests}
    \item \textbf{Obtaining Released Responses}
    \item \textbf{Cleanup} 
\end{itemize}

\subsubsection{Transaction Agent Architecture} $\\$
To cover the whole life-cycle of a transaction, we propose the Transaction Agent architecture in a pipelined manner. As Figure \ref{fig-transaction-agent} shows, the Txn Agent has five main components as follows:
\begin{itemize}
    \item \textbf{Task Loader}. It detects empty transaction entries in on-chip memory. If the transaction entry has been cleaned up and there are transaction workloads pending processing, it will issue AXI memory read operations to load a new txn from the HBM to the FPGA. It will convert the raw txn format to the internal lock list format and reset the counters for this txn entry.
    \item \textbf{Lock-Get Sender}. When a txn has been loaded, it starts sending Lock-Get Requests to lock agents. It also counts the number of locks with data access and puts these locks in a buffer for faster txn commitment.
    \item \textbf{Lock Response Receiver}. We apply a unified interface for the Grant, Abort, Waiting, and Released lock responses. It parses the lock responses from local lock channels according to their response types. It updates the counters and control flags of txns based on the lock responses. In addition, if one txn ends, it can instantly conduct the clean-up workflow when the last lock Released response arrives.
    \item \textbf{Txn Commit Controller}. When all locks of a txn are granted, it will commit the transaction by reading and writing the corresponding data from or to the local HBM. It reads the lock buffer to commit the transaction to reduce unnecessary on-chip memory read. It requires a whole AXI memory channel due to the read and write operations.   
    \item \textbf{Lock-Release Sender}. When the transaction has been committed, it sends the lock release requests to lock agents. As the Lock Agent only has one input lock request, we unify the lock Get and Release requests to share one output port to keep a clean interface. 
\end{itemize}

In addition to the main components, the Txn Agent also has a signal synchronization center with timeout timers. As these components are working on different stages of the transaction life-cycle, they need counters and flags to synchronize. For example, whether a transaction has been loaded or committed to trigger sending lock requests. These components mainly update the counters based on their operations, and the signal synchronization center updates the corresponding flags in one place. This greatly reduces the complexity of signal management. In addition, timeout timers are responsible for counting the time stamp and setting the timeout according to the transaction status.

\subsubsection{Pipelining and Context Switch} $\\$
We split the Txn Agent into five stages to enable pipelining. We implement context switch between transactions to fully utilize the pipeline to improve the transaction throughput. As Figure \ref{fig-one-transaction-pipeline} shows, one transaction pipeline has 6 stages with four stage barriers. The task loading and the lock Get request sending process the transaction sequentially. Only after the txn is fully loaded the Lock-Get Sender will start acquiring locks, which is the first stage barrier. The Transaction Agent accepts all lock responses regardless of the states of Lock-Get Sender and Lock-Release Sender. It is always ready to receive new lock responses once it finishes processing the previous one. Only after all locks are granted, the Txn Commit Controller will starting accessing the local data, which is the second stage barrier. Only after all read operations are finished and local data have been written to the memory, the Lock-Release Sender will release the locks. This is the third stage barrier. Only after all locks are released, the counters and flags will be cleared and the Task Loader will load another transaction to this txn register. Thus, there are four stage barriers in this pipeline. 

As there are multiple Txn Agents in the system, hot locks, lock conflicts, and congestion at the Lock Agent will cause the lock responses to arrive out-of-order. For example, the Lock 0 may get a Waiting response in the beginning and grant later. Lock 2 may have a longer queuing time before entering the Lock Agent and arrive later than others. Note that the time axis in Figure \ref{fig-one-transaction-pipeline} does not represent the real latency. The processing time for each lock is different in these components. Processing time per lock in the Task Loader and Txn Commit Controller are usually longer than others due to the HBM read and write operations (>30 clock cycles). The sending of lock requests and the parsing of lock responses are usually very quick ($\sim$5 clock cycles). However, the latency from sending the request and receiving the lock response depends on the workload and varies greatly.

To make all components busy, we place multiple txn registers in the Txn Agent to let multiple txns run in parallel. Each component of the Txn Agent can dynamically switch to the transaction whose stage is ready to proceed. The Lock Response Receiver directly updates the counters of that transaction based on the lock response information, while other components have a default Context Switch state to iteratively visit all transaction entries to check which is ready. The number of txn registers is a hyper-parameter that needs to be explored for high performance. When there are few txns, the hardware pipeline is underutilized. When there are many txns running in parallel, the utilization of Txn Agent increases, but the total number of active locks also increases, which brings congestion to the crossbars and Lock Agents. Therefore, we should make a utilization-congestion trade-off to achieve high performance.

\subsection{Crossbar}
The crossbar connects the Txn Agents with Lock Agents because the locks of one txn may reside in different lock tables. However, simply connecting them in an all-to-all manner causes challenges to the placement and routing stage when synthesizing the accelerator. This often leads to long wires and long latency, which increases the Worst Negative Slack (WNS) and runtime failures on the FPGA. 

\begin{figure}[t]
    \centering
    \includegraphics[width=\linewidth]{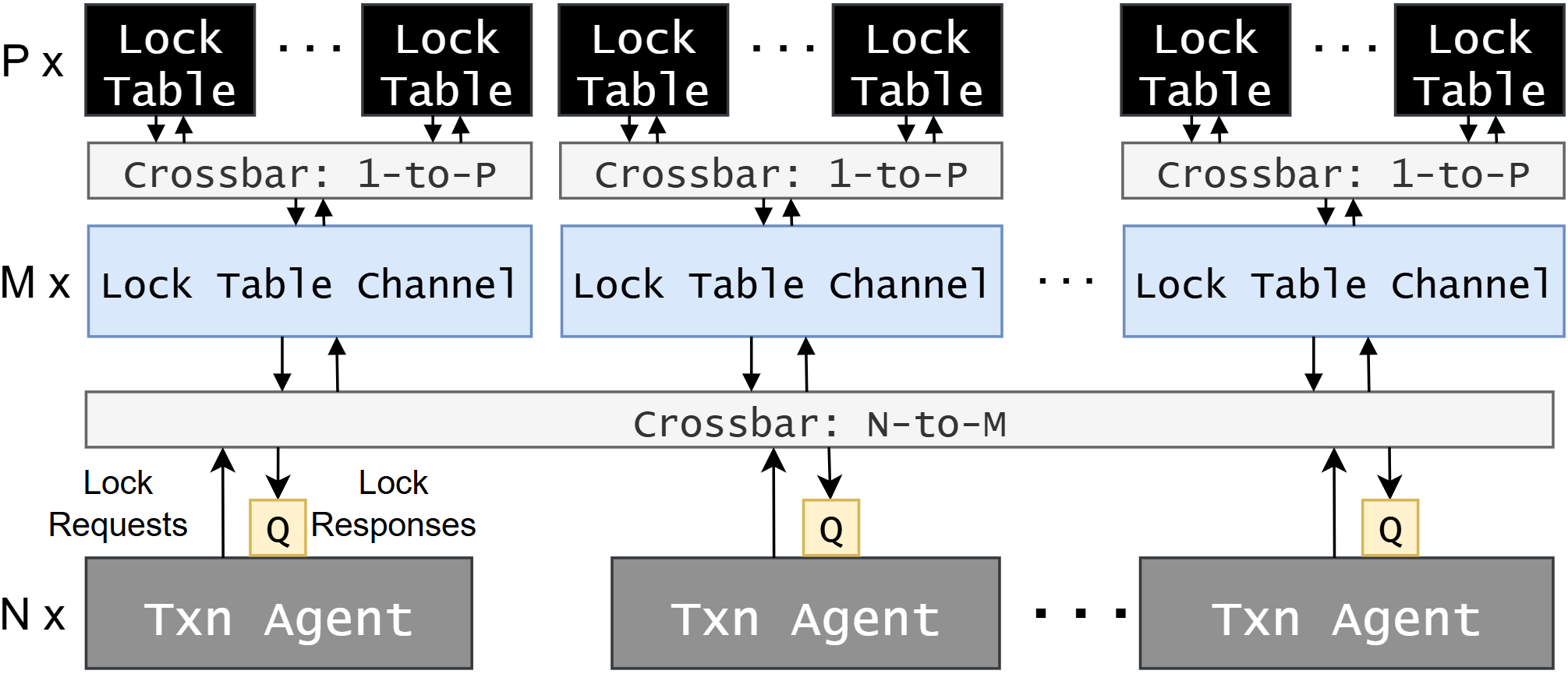}
    \caption{Accelerator architecture with optimized crossbar.}
    \label{fig-crossbar}
\end{figure}

Therefore, we optimize the crossbar by the lock channel-table hierarchy. This can reduce the crossbar complexity while maintaining almost the same performance. As the lowest lock serving time is 3 cycles and the lock requests come somehow randomly, we can cluster these lock tables together to reduce the crossbar size. 

The detailed accelerator architecture with optimized crossbar is presented in Figure \ref{fig-crossbar}. Each Txn Agent has a buffer queue for the lock response to avoid possible stalls on the Lock Agent's side, as such stall will prevent the Lock Agent to accept new lock requests. 
\begin{align}
    Size_{crossbar} ^ {N \times M}  &= N*(1\times M) + M*(N\times1)=2MN \label{eq-NxM-crossbar} \\
    Size_{crossbar} ^ {8 \times 16} &= 8 * (1\times16) + 16*(8\times1) = 256 \label{eq-crossbar-8x16} \\
    Size_{crossbar} ^ {8 \times 4 + 4(1 \times 4)} &= 8 * (1\times4) + 4*(8\times1) + 4*(1 \times 4) = 80 \label{eq-crossbar-4x4x4}
\end{align}
Suppose that there are $N$ Txn Agents and $M * P$ Lock Tables. The size of the crossbar will be $N \times (P * M)$ in all-to-all manner, but $N \times M + P(M \times 1)$ after clustering. Equation \ref{eq-NxM-crossbar} provides a reference complexity for $N \times M$ crossbar, which needs N $1 \times M$ crossbars and M $N \times 1$ crossbars with a total size of 2MN. For example, the $8\times16$ crossbar size for 8 Txn Agents and 16 Lock Tables will be 256 in naive implementation, as Equation \ref{eq-crossbar-8x16} shows. If we apply 4 lock channels and each lock channel has 4 lock agents, then we only need one $8 \times 4$ crossbar and four $1 \times 4$ crossbars with the total size of 80. As Equation \ref{eq-crossbar-4x4x4} shows, the total crossbar size is roughly reduced to 31.25\% of the naive implementation.

\section{Design Space Exploration}
We explore the design space of the accelerator in the simulation environment for efficient configurations of the hyper-parameters. In addition, we need to explore some high-level design principles, e.g., how to balance the number of Txn Agents and Lock Tables, how to balance the parallelism and lock congestion, etc. We profile the latency parameters form the FPGA boards to reflect the real deployment scenario. Thus, we can find the bottlenecks earlier and make optimizations that will be effective on board.

\subsection{Experimental Setup}
We implement the transaction processing accelerator in SpinalHDL 1.13.0 \cite{SpinalHDL} with Scala 2.13.14 and Java 21.0.7 LTS. We use sbt 1.11.3 (runner version: 1.12.0) to generate the Verilog HDL code and Verilator 5.032 \cite{Verilator} to simulate the hardware. We include the memory latency in the simulation setting. Based on the profiling on an AMD Alveo U55C FPGA card, the HBM latency is 288ns from issuing the memory read command to receive the first data chunk. This corresponds to 36 clock cycles of memory latency when the FPGA clock frequency is 125MHz. We refer to the related work \cite{VLDB_2019_2PhaseLocking} to obtain the TPC-C benchmark traces of serializable transaction processing with 64 warehouses. 

We configure the accelerator in this basic setting: single node, 4 lock channels per node, 4 Lock Agents per channel, 4 Txn Agents per node, and 8 concurrent txns per Txn Agent. When we explore one or two dimensions, we keep others the same as the basic setting. Unlike the on-board test that can run tens of thousands of txns in seconds, each Txn Agent conducts 400 txns in the simulation due to time and space constraints. A single simulation usually generates 5-20GB of text-based waveform files, depending on the accelerator configurations. The default timeout for a txn is $2^{13}$ clock cycles. The default maximum size for one txn is 511 locks to cover extra long transactions, although more than 90\% of txns only have no more than 127 locks in the TPC-C benchmark. 


\subsection{Lock Channel and Lock Agent}
We explore the lock channel - lock agent setting in Figure \ref{fig-sim-table-per-channel}. The basic hypothesis behind clustering Lock Tables into Lock Channels is to reduce the crossbar size without affecting the performance. The minimal lock grant latency is 3 clock cycles, thus putting 2-4 tables per lock channel may keep the similar lock serving latency as 1 table per channel while providing higher throughput. 

\subsubsection{Transaction Throughput} 
\begin{figure}[t]
    \centering
    \includegraphics[width=\linewidth]{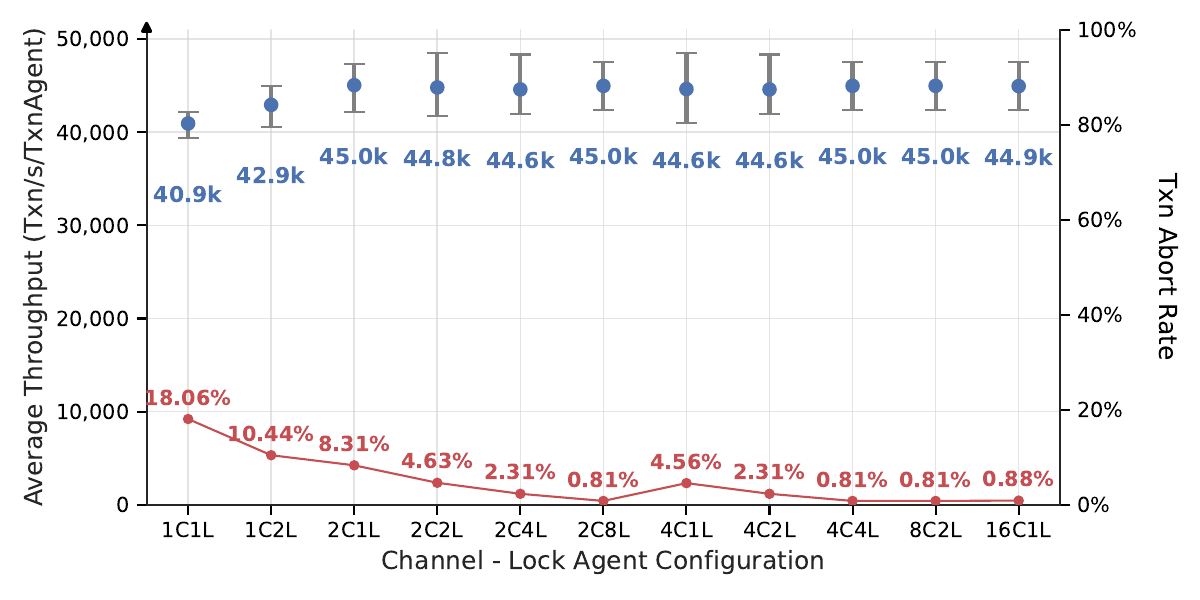}
    \caption{Txn throughput and abort rate across different lock channel - lock agent configurations in simulation.}
    \label{fig-sim-table-per-channel}
\end{figure}

$\\$
We have three findings on how the Lock Channels and Agents configuration benefits the txn throughput according to Figure \ref{fig-sim-table-per-channel}. \textbf{First, the absolute number of Lock Agents matters for good txn throughput.} As the 1T1L (1-channel 1-Lock Agent), 1C2L, and 2C2L settings show, the lock serving efficiency increases with the number of lock agents. The average performance keeps no more than 1\% of difference when there are 4 or more lock agents, regardless of the configuration (e.g., 2 tables with 2, 4, and 8 channels). The 2C4L simulation has 0.89\% lower performance than 1 channel due to the randomness of the workload. 

\textbf{Second, more lock channels should be prioritized over more lock tables if the resources are quite limited.} This can be observed from the difference of the 1C2L and 2C1L configurations. Placing 2 lock agents in 2 channels achieves 4.9\% better performance than placing them into only one channel.

\textbf{Third, the lock agents are really efficient in serving locks}. The 1C1L setting only has 6.9\% lower performance than the 2C2L setting. Serving 4 Txn Agents with 8 txns per agent does not put much pressure on the one lock table. However, if there are more Lock Agents and higher number of parallel txns per agent, we should still choose more than 4 Lock Agents in total to prevent potential bottleneck on the lock serving.


\subsubsection{Transaction Abort Rate} $\\$
Though the Lock Agents provide very efficient lock services, the hot locks requested by many on-going transactions still put pressure on the abort rate. \textbf{The transaction are aborted due to timeouts, not denied locks}. A detailed analysis into the Lock Agents shows that the waiting queues (4K waitQ entries per 64K lock table entries) are long enough to prevent frequent lock denial but transactions still need to wait for conflicting locks to be released. 

\textbf{The key factor to reduce the abort rate is using larger lock tables.} The lock tables increase from 1 to 16 from the 1C1L to 2C8L settings, and the transaction abort rate reduces from 18.06\% to 0.81\%. Similarly, the abort rate reduces from 4.56\% to 0.88\% from 4 tables to 16 tables in the 4C1L and 161L settings. More lock agents can hash the locks to different lock tables to reduce the possibility of false lock conflicts and lock congestion. Therefore, 16 or more lock agents are needed to ensure a low transaction abort rate.


\subsection{Txn Agent and Context Switch}
\begin{figure}[t]
    \centering
    \includegraphics[width=\linewidth]{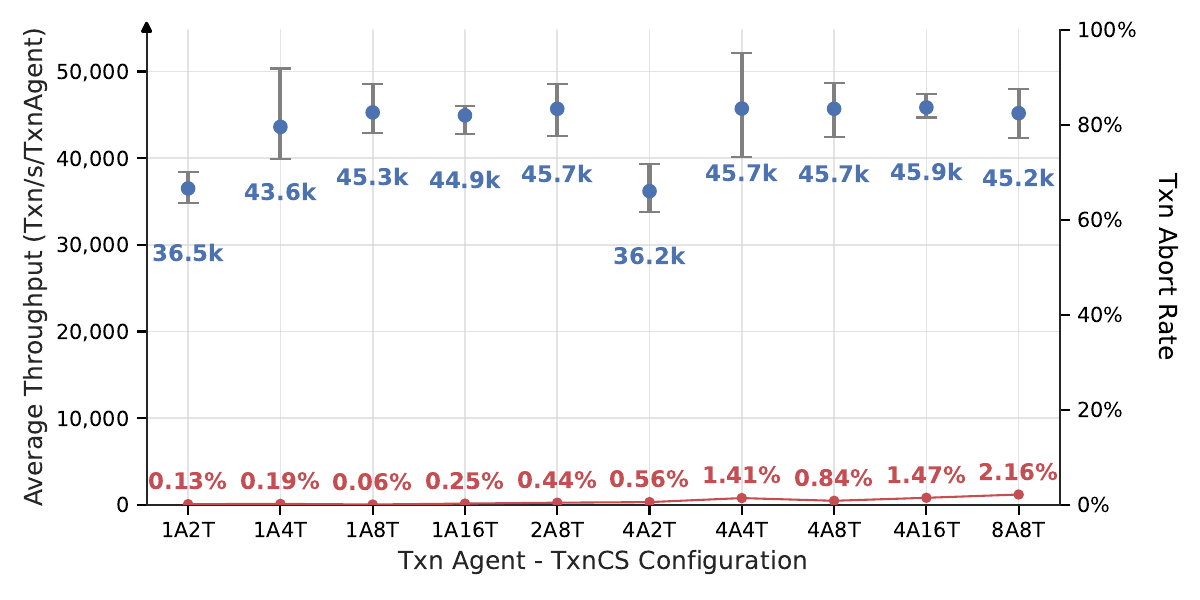}
    \caption{Txn throughput and abort rate across different Txn Agent - Txn Context Switch configurations in simulation.}
    \label{fig-sim-txn-per-agent}
\end{figure}

We explore the Txn Agent - Txns per agent by context switch (TxnCS) in Figure \ref{fig-sim-txn-per-agent}. One Txn Agent conducts several txns at the same time. Thus, the number of Txn Agents and txns per agent together determine the total txn parallelism. We refer to 1 Txn Agent with 2 parallel txns by context switch as 1A2T. 

\subsubsection{Transaction Throughput} $\\$
The transaction throughput is mainly contributed by the number of Txn Agents, then the TxnCS per agent. We summarize the three findings from Figure \ref{fig-sim-txn-per-agent}.
\textbf{First, more TxnCS can bring higher pipeline parallelism to benefit the performance}. As the 1-Agent cases show, increasing the TxnCS from 2 to 4 increases the average txn throughput by 19.46\%, and 8 TxnCS has 3.82\% higher throughput than 4 TxnCS. This also shows that more txns per agent can hide the lock request latency between txns to make the hardware busy. Thus, when the pipeline is already close to full, increasing TxnCS will keep almost the same saturated performance.


\textbf{Second, more Txn Agents bring higher txn processing capability with linearly increased area cost.} Provided enough parallelism, the proposed Txn Agent achieves similar performance when scaling up. For example, the average throughput values of 1A16T, 2A8T, 4A16T, and 8A8T are within 2\% of difference. The total txn processing throughput increases almost linearly with the number of txn agents, achieving a scaling factor of 2.04$\times$, 2.01$\times$, and 1.97$\times$ when doubling the txn agents from 1 to 8. On the other hand, more Txn Agents requires more hardware resources. The scaling up of Txn Agents is actually bounded by the total resources of the FPGA and the placement\&routing congestion during synthesis. 

\textbf{Third, more txns per agent can provide more stable performance.} As Figure \ref{fig-sim-txn-per-agent} shows, the Min-Max bars of 16 TxnCS are obviously smaller than that of 4 and 8 TxnCS. The absolute values of the Min-Max bars range 10.5-12.0K, 5.7-6.2K, and 2.7-3.2K for TxnCS=4, 8, and 16 respectively. The txn throughput of 16 TxnCS is about 3.7-3.9$\times$ more stable than that of 4 TxnCS. The main reason is that more TxnCS can better saturate the hardware to provide stable performance regardless of the various lengths of transactions.
 

\subsubsection{Transaction Abort Rate} $\\$
We observe a very clear trend of \textbf{the transaction abort rate increasing with the total number of concurrent txns}. For example, the abort rate grows from 0.13\% to 2.16\% from 1A2T to 8A8T settings. The main reason is that the total number of locks on-the-fly is nearly linear to the total number of txns. Thus, more concurrent txns will increase the congestion rate of lock requests and the number of hot locks, which produces more lock Waiting responses to cause more txns to be timeout. This trend is observed because we fix the number of lock channels and lock agents per channel to 4. Increasing the total number of lock agents with the number of concurrent txns can mitigate this problem.

\section{Evaluation}
We evaluate our proposed txn processing accelerator on an FPGA board and benchmark it against a CPU baseline.


\subsection{Experimental Setup}
We built the accelerator with Coyote V2 0.2.1 \cite{Coyote_OSDI_2020, CoyoteV2_SOSP_2025} as the FPGA shell in Vivado 2024.2. The FPGA board is AMD/Xilinx Alveo U55C with 16 GB of HBM. We offload the TPC-C benchmark trances with serializable concurrency control on 64 warehouses from CPU to the accelerator by the control software, configure the memory addresses, then start the execution and wait for the finish signal. 

The CPU baseline is based on open-source code from \cite{VLDB_2019_2PhaseLocking}. This baseline is very strong because the C++ implementation can provide better performance than other open-source implementations in Java. The CPU is AMD EPYC 7302P 16-Core/32-Thread @3.0 GHz with 64 GB of memory. We repeat the CPU benchmarking on the same computing node that hosts the FPGA board to ensure the same Operating System (OS) environment. The OS is Ubuntu 22.04.5 LTS with Linux kernel 6.8.0 and GCC 11.4.0. 

\begin{table}[b]
\caption{Resource utilization on AMD/Xilinx U55C FPGA.}
\label{tab-fpga-utilization}
\small
\begin{tabular}{c|r|r|r|r|c|r|r}
\toprule
\hline
     & \multicolumn{1}{c|}{\begin{tabular}[c]{@{}c@{}}LUT\\ Logic\end{tabular}} & \multicolumn{1}{c|}{\begin{tabular}[c]{@{}c@{}}LUT\\ Mem.\end{tabular}} & \multicolumn{1}{c|}{\begin{tabular}[c]{@{}c@{}}CLB\\ Reg.\end{tabular}} & \multicolumn{1}{c|}{\begin{tabular}[c]{@{}c@{}}Block\\ RAM\end{tabular}} & \begin{tabular}[c]{@{}c@{}}AXI\\ Ch.\end{tabular} & \multicolumn{1}{c|}{\begin{tabular}[c]{@{}c@{}}Clock\\ Cycle\end{tabular}} & \multicolumn{1}{c}{WNS} \\ \hline
2A2T & 22.54\%                                                                  & 22.27\%                                                                 & 17.41\%                                                                 & 15.97\%                                                                  & 4                                                 & 4ns                                                                       & -0.53ns                   \\ \hline
4A8T & 24.64\%                                                                  & 22.67\%                                                                 & 19.07\%                                                                 & 19.00\%                                                                 & 8                                                 & 4ns                                                                        & -0.78ns                   \\ \hline
4C2L & 26.97\%                                                                  & 40.22\%                                                                 & 19.30\%                                                                 & 19.20\%                                                                 & 8                                                 & 5ns                                                                        & -0.57ns                   \\ \hline
\bottomrule
\end{tabular}
\end{table}

\subsection{Resource Utilization}
We synthesized three configurations on the FPGA successfully. The first configuration has 2 lock channels, 2 Lock Agents per channel, 2 Txn Agents, and TxnCS=2 (2C2L-2A2T). The other configurations are 2C2L-4A8T and 4C2L-4A8T. The resource utilization and timing results are listed in Table \ref{tab-fpga-utilization}. Two more Txn Agents occupy 1.9\% of LUT logic, 0.4\% of LUT memory, 1.7\% of CLB register, and 3.0\% of Block RAM. As each AXI channel has one Read and one Write bus, the 4 more 512-bit AXI channels result in 2048 bits of more data lines. However, the 17.5\% more LUT memory of 4 lock channels brings challenges to the placement and routing. We apply a 25\% longer clock cycle to maintain a usable Worst Negative Slack (WNS). The means that the design has challenges during synthesis due to its scale and complexity. Longer WNS makes the design more unstable under high temperature. Therefore, we should further optimize synthesis, placement, and routing for larger configurations. 

\subsection{Benchmarking Results}
\begin{figure}[t]
    \centering
    \includegraphics[width=\linewidth]{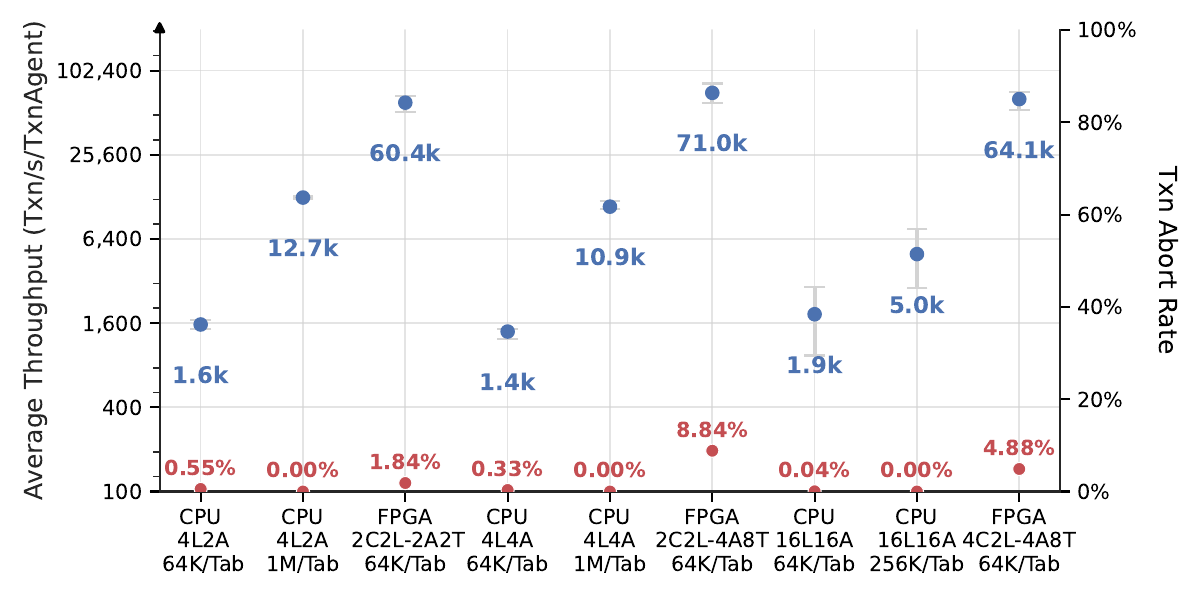}
    \caption{Txn throughput and abort rate of TPC-C benchmark on CPU and FPGA in the same computing node.}
    \label{fig-cpu-fpga-tpcc}
\end{figure}

We present the performance per agent and transaction abort rate in Figure \ref{fig-cpu-fpga-tpcc} and the total performance in Table \ref{tab-cpu-fpga-tpcc}. We provide three cases of CPU baselines to provide a fair comparison: First, the CPU has the same number of lock agents, transaction agents, and lock table size (CPU-4L2A-64K/Tab and CPU-4L4A-64K/Tab). Second, the CPU has the same threads but 16$\times$ larger lock table size, considering the huge advantage of the 64GB of CPU memory (CPU-4L2A-1M/Tab and CPU-4L4A-1M/Tab). Third, the CPU launches all the threads with the same lock table size per agent or even larger table sizes (CPU-16L16A-64K/Tab and CPU-16L16A-256K/Tab). 

As Figure \ref{fig-cpu-fpga-tpcc} shows, our FPGA accelerator is 38.6$\times$ and 50.9$\times$ faster than the CPU baseline with the same lock table size and number of agents. Provided enough lock table size, the CPU baseline achieves zero abort rate but is still 4.8$\times$ and 6.5$\times$ slower than the FPGA. Our FPGA accelerator scales up better than CPU, but scaling up the FPGA and CPU both have performance drop in the per agent performance. The main reason for the FPGA performance loss is due to 25\% slower clock cycle and the crossbar complexity that limits the number of lock agents. The performance drop on CPU is mainly due to lock congestion, which is also reflected in the Min-Max bars of the average throughput.  

Our 4 Lock Agents achieve a throughput of 6.3-6.7M lock/s in the 2A2T setting and 15.8-16.7M lock/s in the 4A8T setting, which is proportional to the transaction throughput. The CPU baseline has 0.2-0.3M lock/s with the same table size, 1.5-2.5M lock/s with 1M lock/table and 4 lock agents, and 1.6-6.5M lock/s with 16 agents. The lock serving of our proposed Lock Agent is 35.2-52.1$\times$ faster than the CPU threads with the same lock table size, and is still 4.4-6.7$\times$ faster than the CPU threads with 16$\times$ larger lock tables. Our accelerator achieves 2.5$\times$ higher lock serving throughput even compared to the best case of 16-thread CPU.  

\begin{table}[b]
\caption{Txn throughput of TPC-C benchmark on CPU and FPGA in the same computing node.}
\label{tab-cpu-fpga-tpcc}
\begin{tabular}{l|r|c|c|r|r}
\toprule
\hline
Device   & \begin{tabular}[c]{@{}l@{}}Lock/\\ Agent\end{tabular} & \begin{tabular}[c]{@{}c@{}}Lock \\ Agents\end{tabular} & \begin{tabular}[c]{@{}c@{}}Txn\\ Agents\end{tabular} & \multicolumn{1}{l|}{\begin{tabular}[c]{@{}l@{}}Total \\ Txn/s\end{tabular}} & Speedup \\ \hline
CPU-Same & 64K                                                   & 4                                                      & 2                                                    & 3130                                                                        & 1.0 $\times$    \\ \hline
CPU-Huge & 1024K                                                 & 4                                                     & 2                                                    & 25313                                                                       & 8.1  $\times$     \\ \hline
2C2L-2A2T     & 64K                                                   & 4                                                     & 2                                                    & 120840                                                                      & 38.6 $\times$    \\ \hline \hline
CPU-Same & 64K                                                   & 4                                                      & 4                                                    & 5572                                                                        & 1.0  $\times$     \\ \hline
CPU-Huge & 1024K                                                 & 4                                                      & 4                                                    & 43582                                                                       & 7.8  $\times$     \\ \hline
2C2L-4A8T     & 64K                                                   & 4                                                      & 4                                                    & 283835                                                                      & 50.9 $\times$     \\ \hline
CPU-Big  & 64K                                                   & 16                                                     & 16                                                   & 29653	                                                                        & 5.3 $\times$      \\ \hline
CPU-Huge & 256K                                                  & 16                                                     & 16                                                   & 79818                                                                      & 14.3  $\times$    \\ \hline
4C2L-4A8T     & 64K                                                   & 8                                                      & 4                                                    & 256261                                                                      & 46.0 $\times$     \\ \hline
\bottomrule
\end{tabular}
\end{table}

As Table \ref{tab-cpu-fpga-tpcc} shows, compared to the best cases of CPU in 32 threads with 4$\times$ and 16$\times$ larger total lock table size, our FPGA implementation with just 4 lock agents and 4 txn agents is 3.6-9.6$\times$ faster. Note that the zero abort rate of the CPU-16L16A-256K/Tab shows that the lock table size is not constraining its performance. 

\section{Conclusion}
In this paper, we propose an FPGA-based transaction processing accelerator. The Lock Agent is optimized for low-latency lock granting and releasing. The Transaction Agent is optimized with asynchronous pipelines for high transaction processing parallelism. And we specially optimize the crossbar to reduce the hardware complexity and synthesis cost. We conduct a design space exploration on the architecture to find design insights and efficient configurations. The evaluation results show that our accelerator achieves 35.2-52.1$\times$ higher lock serving throughput and 38.6-50.9$\times$ higher transaction throughput than the CPU baseline with the same configuration. We achieve 283K Transaction/s on a single node with around 30\% of FPGA resource utilization, which is 3.6-9.6$\times$ faster than the CPU applying all 32 threads for transaction processing. 

In the future, we will further optimize the accelerator architecture and the synthesis stage for better scaling-up performance. We will extend it to multiple computing nodes for distributed transaction processing. This transaction processing system can also serve as an experimental platform to study how new cross-node interconnects like CXL and Unified Bus benefit database applications, including the simulation and on-board testing environments.

\begin{acks}
This project is supported by ETH Future Computing Lab.
\end{acks}

\bibliographystyle{ACM-Reference-Format}
\bibliography{reference}

\appendix

\end{document}